\documentclass[10pt,letterpaper]{article}

\usepackage{cogsci}
\usepackage{hyperref}
\usepackage{amsmath}
\usepackage{amssymb}
\usepackage{booktabs}

\cogscifinalcopy 

\usepackage{pslatex}
\usepackage{apacite}
\usepackage[dvipsnames]{xcolor}
\usepackage{colortbl}

\newcommand{\ci}[3]{#1; 95\%~CI~=~[#2,~#3]}
\newcommand{\chisq}[3]{$\chi^{2}(\text{#1})$~=~#2, $p$~#3}

\usepackage{float} 
\usepackage{subfigure}
\usepackage{natbib}
\usepackage{booktabs}
\usepackage{graphicx}

\title{Measuring and predicting variation in the difficulty of questions about data visualizations}

\author{{\large \bf Arnav Verma (arnavv@stanford.edu)} \\
  Department of Psychology, Stanford University \\
  Stanford, CA, United States
  \AND {\large \bf Judith E. Fan (jefan@stanford.edu)} \\
  Department of Psychology, Stanford University \\
  Stanford, CA, United States
}

\begin{document}
\definecolor{wainer}{HTML}{3a0ca3}
\newcommand{\wanier}{\textcolor{wainer}{\textbf{WAN}}}
\definecolor{ggr}{HTML}{e26d5c}
\newcommand{\ggr}{\textcolor{ggr}{\textbf{GGR}}}
\definecolor{brbf}{HTML}{f4a261}
\newcommand{\brbf}{\textcolor{brbf}{\textbf{BRBF}}}
\definecolor{vlat}{HTML}{247ba0}
\newcommand{\vlat}{\textcolor{vlat}{\textbf{VLAT}}}
\definecolor{calvi}{HTML}{70c1b3}
\newcommand{\calvi}{\textcolor{calvi}{\textbf{CALVI}}}

\maketitle
\begin{abstract}

Understanding what is communicated by data visualizations is a critical component of scientific literacy in the modern era.
However, it remains unclear why some tasks involving data visualizations are more difficult than others.
Here we administered a composite test composed of five widely used tests of data visualization literacy to a large sample of U.S. adults ($N=503$ participants).
We found that items in the composite test spanned the full range of possible difficulty levels, and that our estimates of item-level difficulty were highly reliable. 
However, the type of data visualization shown and the type of task involved only explained a modest amount of variation in performance across items, relative to the reliability of the estimates we obtained.
These results highlight the need for finer-grained ways of characterizing these items that predict the reliable variation in difficulty measured in this study, and that generalize to other tests of data visualization understanding.




\end{abstract}

\section{Introduction}
\begin{figure*}[ht!]
    \centering
    \includegraphics[width=1\linewidth]{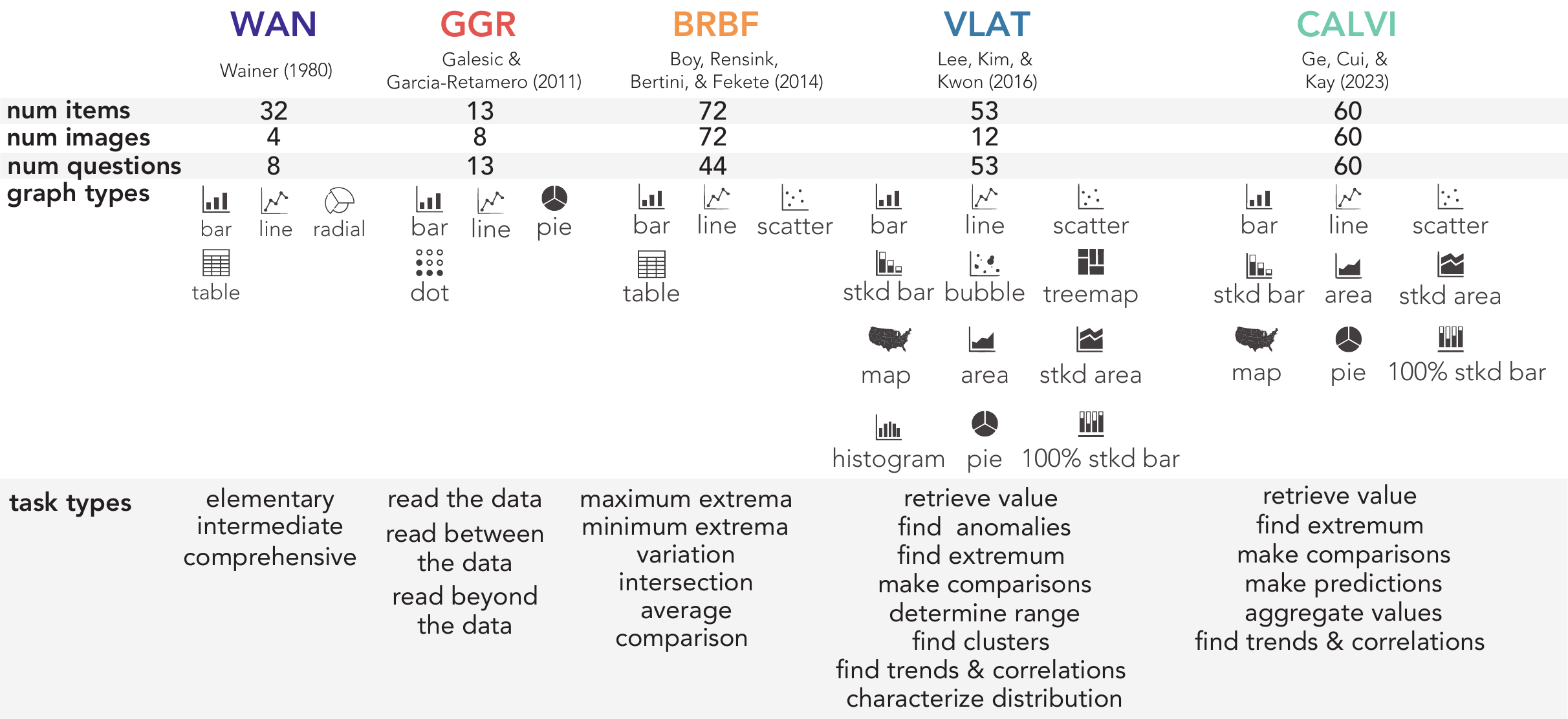}
    \caption{We used 230 items from five popular tests of data visualization literacy, which vary in \emph{graph type} and \emph{task type}.}
    \label{fig:test_overview}
\end{figure*}

Data visualizations are powerful tools invented by humans for making sense of a complex world.
Although they have only existed for a few centuries, they are practically indispensable in modern scientific workflows \citep{munzner2014visualization} and are pervasive in social media and the news \citep{lee2021viral}. 
Data visualizations (or equivalently, \textit{graphs}, \textit{charts}, or \textit{plots}) enable people to reason about quantitative information through visual encodings of data \citep{keim2008visual, munzner2014visualization}. 
A single visualization can even serve multiple purposes. 
For example, a scatter plot can help with finding outliers in the data while also assisting in deriving broader insights about complex trends \citep{boy2014principled,lee2016vlat, borner2019data, lundgard2021accessible}.


Performing these tasks relies on the coordination of many cognitive processes, including rapid visual computations \citep{cleveland1984graphical, ciccione2023trend, cui2021synergy}, invocation of the appropriate graph schema \citep{pinker2014theory}, mathematical operations \citep{gillan1994componential}, control of finite attentional and working memory resources \citep{padilla2018decision}, and other reasoning processes to derive general insights informed by prior knowledge \citep{carpenter1998model, shah2011bar}. 
However, people do not automatically acquire the ability to reason about data visualizations; rather, this ability is acquired gradually, and usually in formal educational contexts \citep{alper2017visualization, mix2012relation, wainer1980test}. 
However, the effectiveness of educational interventions for helping people develop core data visualization literacy skills remains unclear. 


This lack of clarity reflects, in part, the lack of a coherent suite of reliable and valid tools for measuring data visualization literacy.
Several test-based measures currently exist, each of which generally consist of a series of items, with each consisting of a question paired with a data visualization \citep{delmas2005using,galesic2011graph,boy2014principled,lee2016vlat,pandey2023mini,ge2023calvi,maltese2015data}.
However, because they have not been directly compared, the extent to which they reliably measure the same underlying construct and whether they imply a consistent decomposition of data visualization literacy into distinct components remains unknown \citep{brockbank2025evaluating,borner2019data, brehmer2013multi, friel2001making}.
Some tests contain items meant to measure a compact hierarchy of abstract abilities --- e.g., progressing from ``reading the data'' to ``reading beyond the data'' \citep{galesic2011graph, wainer1980test} --- while other tests are designed to assess performance on a broader suite of more concrete tasks, such as finding extreme values or detecting correlations \citep{lee2016vlat, ge2023calvi, pandey2023mini, boy2014principled}. 
Additionally, there are tests which also focus on measuring the ability to overcome misleadingly constructed data visualizations, such as ones using inappropriate axis limits \citep{ge2023calvi}. 
Here we leverage the diversity of the tasks and data visualizations represented across several existing tests to develop consistent procedures for measuring and comparing the difficulty of tasks involving data visualizations.

In this study, we aggregated 230 items from five tests of data visualization literacy: a 32-item assessment developed by \cite{wainer1980test}, which we refer to as \wanier{}; a 13-item assessment developed by \cite{galesic2011graph}, which we refer to as \ggr{}; a 72-item assessment developed by \cite{boy2014principled}, which we refer to as \brbf{}; a 53-item Visualization Literacy Assessment Test (\vlat{}; \cite{lee2016vlat}); and a 60-item assessment known as \calvi{} \citep{ge2023calvi}. 
Together, these assessments represent some of the most widely used and influential tools for measuring data visualization literacy in several research communities, including computer science, education, and psychology.


\section{Method}
\vspace{-0.5em}
\subsection{Participants}
We used Prolific to recruit 503 U.S.-based participants who spoke English as their primary language and have maintained an approval rate of at least 95\%. We maximized the number of participants recruited with the resources available, allowing us to obtain at least 80 responses for each item in the stimuli set. 

All participants were given up to three opportunities to complete the warmup trials that included items from the National Assessment of Educational Progress intended to assess middle-school level quantitative literacy skills. Participants who did not pass after three attempts on these tasks did not proceed to the main experiment.

In total, there were 37 participants who failed to complete the warm up trials and additionally another 40 participants who quit the study before completing at least 50\% of all trials (23 items). These 77 participants were all omitted from our analysis, leading to a total of 426 participants being used in our analysis and a range of 80 to 92 responses per item.

\subsection{Materials}
We aggregated 230 multiple-choice items from five widely used assessments of data visualization understanding (Figure \ref{fig:test_overview}): \wanier{} \citep{wainer1980test}, \ggr{} \citep{galesic2011graph}, \brbf{} \citep{boy2014principled}, \vlat{} \citep{lee2016vlat}, \calvi{} \citep{ge2023calvi}.

All assessments categorized items using at least two common features: by \emph{task type}, which refers to the reasoning steps a participant performs to answer a question, and by \emph{graph type}, which describes how the image encodes data into visual features. 

Since different assessments sometimes use different labels for similar tasks, we additionally defined a simpler common set of \emph{task types} that could apply to all assessments: \texttt{value identification}, where participants need to retrieve an individual value appearing in a plot (e.g., finding the maximum value); \texttt{arithmetic computation}, where participants are expected to perform arithmetic operations over multiple values displayed in the plot (e.g., finding the average of two values); and \texttt{statistical inference}, where participants are required to estimate latent parameters based on the values shown (e.g., judge the strength of trends or presence of clusters).

To explore the potential impact of presenting information in a data visualization as opposed to a table, we also included a small number of table-based items that were otherwise equivalent to the visualization-based ones.


\begin{figure*}[ht!]
    \centering
    \includegraphics[width=0.75\linewidth]{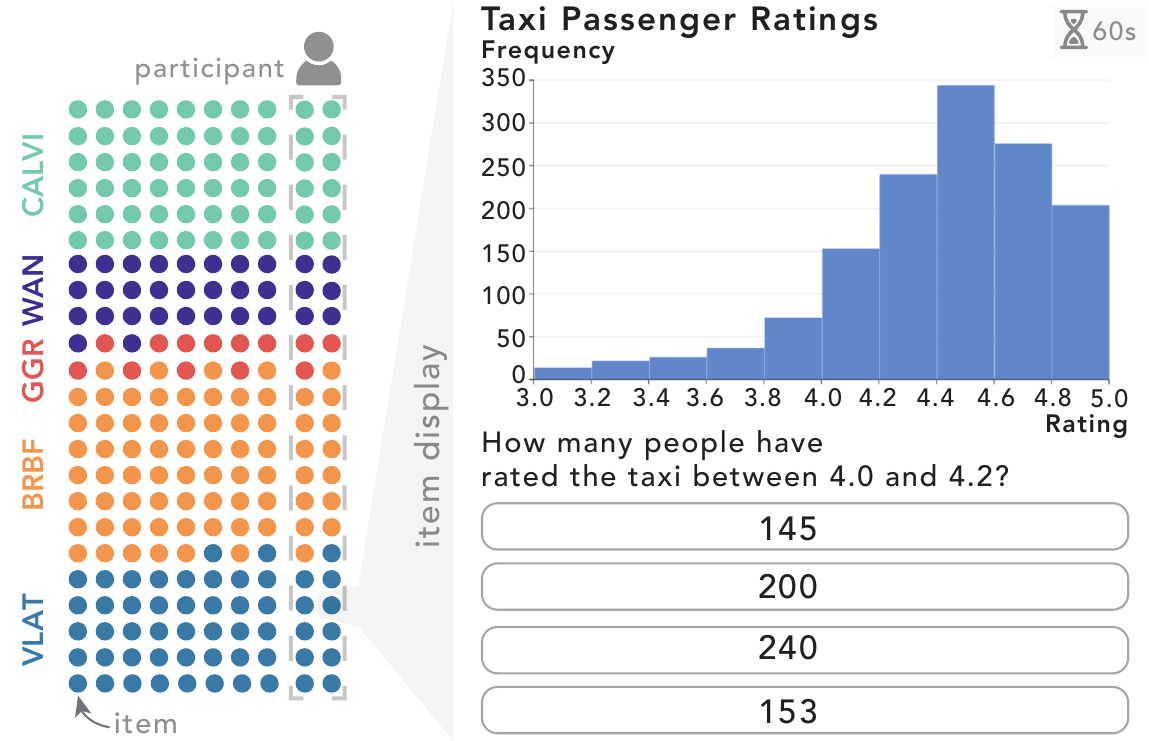}
    \caption{Our item sampling procedure selected 46 items from the total set of 230 items for each participant (left). Everyone was presented with multiple-choice items from all six tests, with a 60 second time limit to answer each question (right).}
    \label{fig:procedure_overview}
\end{figure*}

\textbf{\wanier{}} The test by \cite{wainer1980test} was developed to evaluate children at the third- to fifth-grade level in the United States and includes 32 items. It uses six questions which are paired across one table and three images with different \emph{graph types}: \texttt{line chart}, \texttt{bar chart}, and \texttt{radial plot}.

\textbf{\ggr{}} The test developed by \cite{galesic2011graph} is a widely used 13-item assessment comprising of three \texttt{bar plots}, three \texttt{line plots}, an \texttt{icon array}, and a \texttt{pie chart}. It was initially designed to explore a compact hierarchy of abstract abilities, progressing from ``reading the data" to ``reading between the data" and finally, ``reading beyond the data." Originally, nine of the test items required a numerical response, and four were multiple choice. However, to maintain consistency with other test items, we mapped items requiring a numerical response into a multiple-choice format by selecting the top four most frequent responses based on a prior study \citep{verma2024evaluating}.

\textbf{\brbf{}} The test by \cite{boy2014principled} measures the influence of different data and visual properties used across three different \emph{graph types}. It originally consisted of 60 \texttt{bar charts}, 60 \texttt{scatter plots}, 120 \texttt{line graphs}, and 48 \texttt{tables}. All items were initially categorized by six \emph{task types}: maximum extrema, minimum extrema, intersection, variation, average, and comparison. Each combination of \emph{task type} and \emph{graph type} included at least two unique questions, five images of charts, and one image of a table. We create a subset of 72 items representing each unique question, \emph{task type}, and \emph{graph type} combination used in the assessment to reduce the total number of items while maintaining item diversity across different categories.

\textbf{\vlat{}} The Visualization Literacy Assessment Test by \cite{lee2016vlat} is an influential 53-item assessment containing 12 chart images generated from real-world data sources with a unique image of a \texttt{line chart}, a \texttt{bar chart}, a \texttt{stacked bar chart}, a \texttt{normalized stacked bar chart}, a \texttt{pie chart}, a \texttt{histogram}, a \texttt{scatter plot}, a \texttt{bubble chart}, an \texttt{area chart}, a \texttt{stacked area chart}, a \texttt{choropleth map}, and a \texttt{tree map}. To maintain consistency with other assessments items, we re-classify the \texttt{bubble chart} to a \texttt{scatter plot}. The test originally grouped questions into eight \emph{task types}: retrieving values from a graph, finding correlations \& trends, finding anomalies, finding extrema, making comparisons between values, characterizing distributions, determining the range of values in a graph, and finding clusters of common values.

\textbf{\calvi{}} The Critical Thinking Assessment for Literacy in Visualizations by \cite{ge2023calvi} is a 60-item test that contains 45 items intended to mislead users with unconventional graphs and questions, alongside 15 standard items, following the design guidelines of \vlat{}.
This includes a subset of \emph{graph types}: \texttt{line chart}, \texttt{bar chart}, \texttt{stacked bar chart}, \texttt{normalized stacked bar chart}, \texttt{pie chart}, \texttt{scatter plot}, \texttt{area chart}, \texttt{stacked area chart}, and \texttt{choropleth map}; and a subset of original \emph{task types}: retrieve value, find trends \& correlations, find extremum, make comparisons, make predictions, aggregate values.

\begin{figure*}[ht!]
    \centering \includegraphics[width=0.95\linewidth]{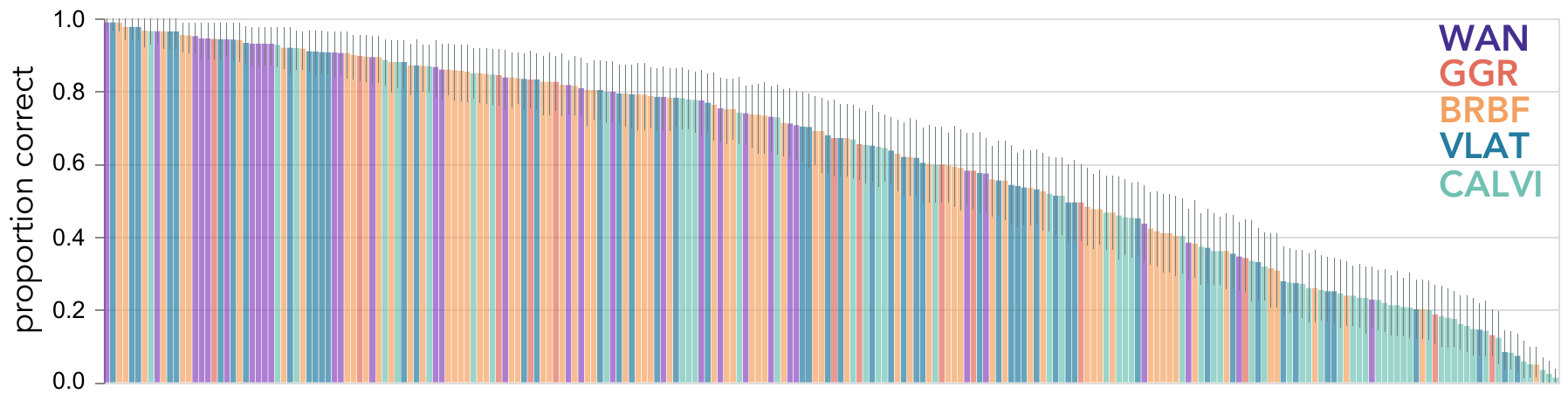}
    \caption{Average performance across all items. Items belonging to the same \emph{test} share the same color. Error bars represent bootstrapped 95\% confidence intervals.}
    \label{fig:item_performance}
\end{figure*}

\subsection{Procedure}
We evaluated all participants on a representative subset of items across all five tests (Figure~\ref{fig:procedure_overview}). 
Specifically, each participant completed 46 items (20\% of the total stimulus set) containing items sampled evenly from all five tests, \emph{graph types}, and \emph{task types}.
To ensure that all participants already possessed basic quantitative literacy skills, each session began with five questions taken from the version of the National Assessment of Educational Progress (NAEP) assessment administered to fourth-grade students.
Items from the same assessment were presented within the same block. Participants were given a maximum of 60 seconds to answer each question, and provided with immediate feedback indicating whether each response was correct. 


\vspace{-0.5em}
\section{Results}
\paragraph{To what degree do items vary in difficulty?}

To determine if there was reliable variation in average performance within our selected test items, we assessed the degree to which the full set of items, pooled across assessments, spanned a wide range of difficulty levels (Figure \ref{fig:item_performance}).
Our results suggest that this suite of assessments cover nearly the entire range of possible levels of difficulty, ranging from items that nearly all participants succeeded on (max. proportion correct: \ci{0.99}{0.96}{1.00}) to items that nearly all participants responded to incorrectly (min. proportion correct: \ci{0.012}{0.0}{0.036}). 
Moreover, the average gap in performance \textit{between} items reliably exceeded the degree of uncertainty in estimates of performance on individual items (lowest-precision estimate: \ci{0.46}{0.34}{0.57}), indicating that the observed variation in performance across items is reliable. 
We also found that participants performed below chance across 36 items out of the 230 total items (proportion correct furthest from chance: \ci{0.012}{0.0}{0.036}).

\vspace{-1em}
\paragraph{How does performance vary across \emph{tests}?}
While some of these tests were designed to measure data visualization literacy in the general adult population, others, such as \wanier{}, aimed to assess data visualization comprehension skills among elementary-school-aged children, and \calvi{} focuses on the skills needed to detect misleading graphs. However, these tests have never before been directly compared to each other under consistent testing conditions.

\begin{figure*}[ht!]
    \centering \includegraphics[width=0.95\linewidth]{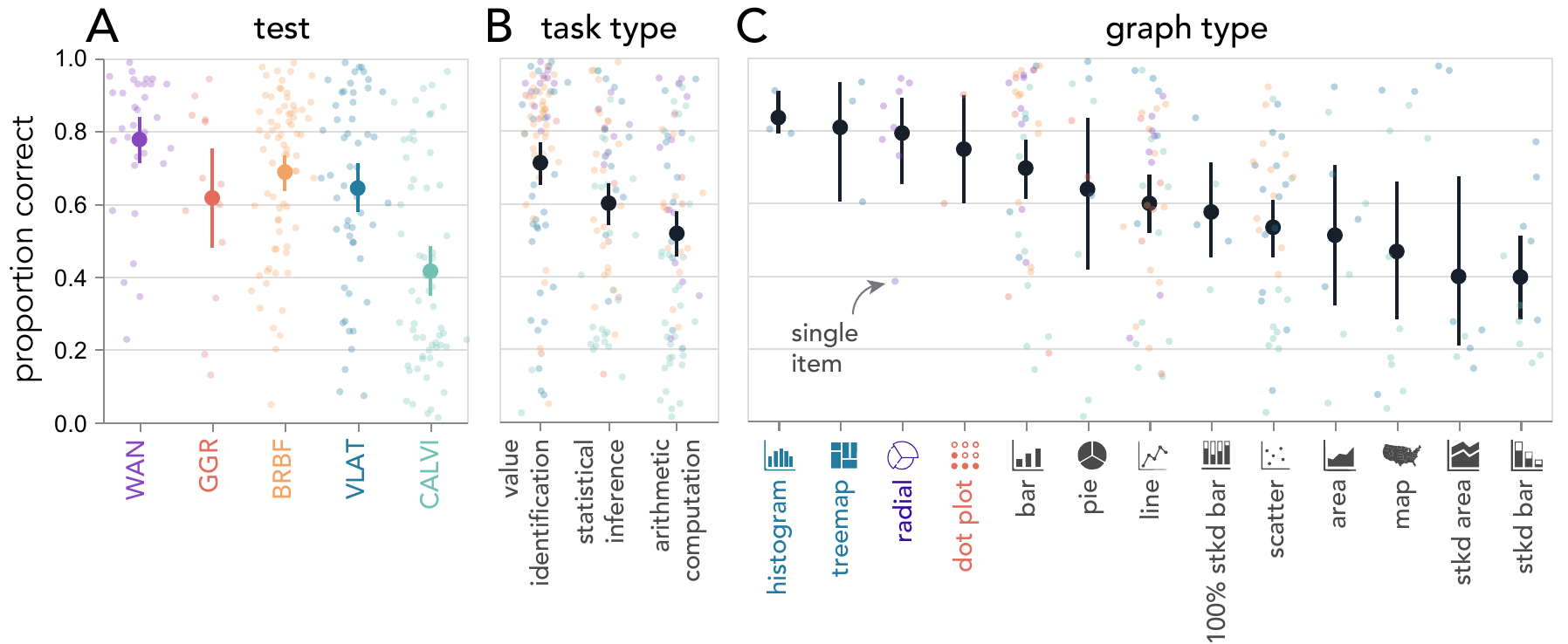}
    \caption{Performance across different \emph{tests} (A), \emph{task types} (B), and \emph{graph types} (C), measured by the mean proportion of correct responses. Opaque dots indicate the mean proportion of correct responses for individual items. Error bars represent bootstrapped 95\% confidence intervals.}
    \label{fig:feature_performance}
\end{figure*}

Here we compared differences in performance across all five tests (Figure \ref{fig:feature_performance}A) by fitting a logistic mixed-effects regression model predicting success on each trial, with \emph{test} as a fixed-effects predictor and random intercepts for different items.
We found that performance differs significantly between tests (\chisq{4}{39.764}{$<$~.001}), and on average, the suite of tests together covers a range of difficulties. 
This includes \wanier{}, where participants' performance is closest to the ceiling (\ci{0.78}{0.71}{0.84}), and \calvi{} (\ci{0.41}{0.35}{0.48}), where participants performed closest to chance (\calvi{} average chance: 0.28), with other tests varying between these (\ggr{}: \ci{0.62}{0.48}{0.75}; \brbf{}: \ci{0.69}{0.63}{0.73}; \vlat{}: \ci{0.64}{0.58}{0.71}).

Taken together, these results are consistent with the notion that some of these tests are reliably harder than others, perhaps because they probe more advanced skills.

\paragraph{How does performance vary across \emph{task types}?}
Perhaps one of the most salient ways different items can differ is the type of task they require people to perform, with some tasks being relatively simple (e.g., retrieving a single value from the visualization) and other tasks requiring additional computation (e.g., inferring the correlation between two variables).
Here we sought to evaluate the degree to which there was a reliable difference in performance across items belonging to the three different categories of tasks: \texttt{value identification}, \texttt{arithmetic computation}, \texttt{statistical inference} (Figure~\ref{fig:feature_performance}B).
As before, we fit a logistic mixed-effects regression model predicting success on each trial from \emph{task} as a fixed-effects predictor and random intercepts for different items.
This analysis revealed reliably different levels of performance across \emph{task types}
(\chisq{2}{13.847}{$<$~.001}; \texttt{value identification}: \ci{0.71}{0.65}{0.77}; \texttt{statistical inference}: \ci{0.60}{0.54}{0.65}; \texttt{arithmetic computation}: \ci{0.51}{0.45}{0.58}). 
These findings are compatible with multiple possibilities, including the notion that some tasks are inherently harder than others --- for instance, that \texttt{value identification} is easier than \texttt{arithmetic computation} with \texttt{statistical inference} in between. 
However, they also remain consistent with the possibility that these differences are a product of the way that the specific questions were posed (or the response options generated) in these tests, such that it is possible, in principle, for \texttt{value identification} to be arbitrarily difficult or to design easier \texttt{arithmetic computation} items that reduce the gaps between them. 



\paragraph{How does performance vary across different \emph{graph types}?} 

Another prominent feature by which these items differ is the visual encoding used to present data.
For example, a \texttt{bar chart} maps numerical values to the height of bars, while a \texttt{stacked bar chart} not only uses bar height to convey quantitative values, but can additionally convey the distribution of values of a categorical variable. 
As such, it might be the case that some types of visualizations are simply more complex than others, and thus more difficult to comprehend. 
Here, we evaluate the degree to which there was a reliable difference in performance across the 13 different types of data visualizations (Figure~\ref{fig:feature_performance}C). 
As in the last section, we fit a logistic mixed-effects regression model predicting success on each trial from \emph{graph type}, with random intercepts for different items, excluding those with tables.

We found that items using different \emph{graph types} reliably differed in performance (\chisq{12}{29.106}{~=~.004}), with the most difficult being \texttt{stacked bar charts} (\ci{0.40}{0.28}{0.51}) and the easiest being \texttt{histograms} (\ci{0.83}{0.79}{0.91}).
However, some \emph{graph types}, like \texttt{tree maps}, only appeared only in a single test, so performance on them is difficult to interpret, as it may rely on the difficulty of the test it came from rather than the type of graph. 
These findings provide some support for the notion that some types of graphs are inherently easier to interpret than others, though further work that uses more controlled manipulations of \emph{graph type} (avoiding confounds with test) would provide a stronger test of this possibility. 


\begin{figure}[ht!]
    \centering 
    \includegraphics[width=0.90\columnwidth]{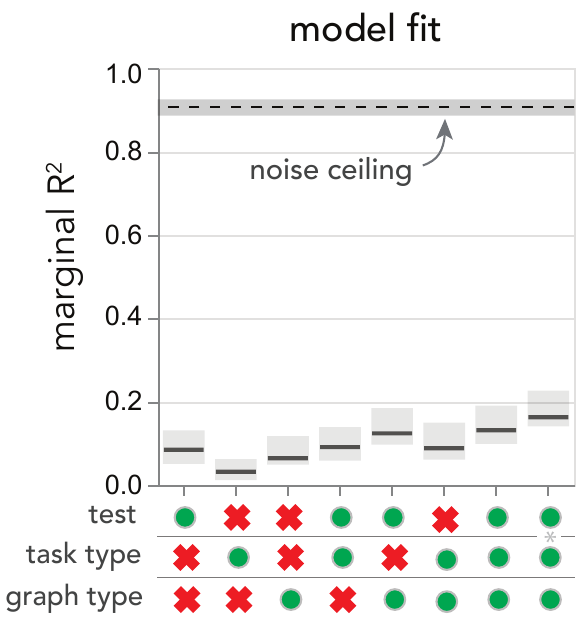}
    \caption{Comparison of model fit across mixed-effects logistic regression models, measured using marginal $R^2$. Green circles indicate fixed effects included in the model, with `*' indicating an interaction between fixed effects. The noise ceiling is estimated by computing the squared Pearson correlation between split halves over participants' data. Gray bands represent the expected variation $R^2$ due to sampling variability across samples of participants, estimated by bootstrap resampling.}
    \label{fig:model_comparison}
    \vspace{-1.5em}
\end{figure}

\vspace{-1em}
\paragraph{How well do all of these features explain variation in performance?}
So far we have found that all three features explain some amount of variation in item-level difficulty.
To what degree do they account for unique or shared variance?
To answer that question, we examined to what degree every combination of these features further improved fit to our performance data. 
We fit every combination of these three features as fixed effects to different mixed-effects logistic regression models to participant errors (7 total; Figure~\ref{fig:model_comparison}). 
We additionally fit a model that includes an interaction term between \emph{test} and \emph{task type}, reflecting the possibility that the difficulty of an \texttt{arithmetic computation} item might depend on which test it was sourced from. Since not all \emph{graph types} were paired with all three \emph{task types}, we omitted the interaction between \emph{graph type} and \emph{task type}. 
Variation across items was modeled as a random effect.

We found that while the model with only \emph{task type} as a fixed effect predicted the least amount of variance in the data (marginal $R^2$: \ci{0.03}{0.01}{0.06}), model fit was improved significantly with the addition of \emph{test} as a fixed effect (\chisq{4}{29.233}{~$<$~.001}), and moderately with \emph{graph type} as a fixed effect (\chisq{12}{27.387}{~=~.006}).

We found that including all three factors provided the best fit (marginal $R^2$: \ci{0.13}{0.09}{0.19}), with a reliable interaction between test and task (marginal $R^2$: \ci{0.16}{0.14}{0.23}).
Nevertheless, we found that all models still fell short of the noise ceiling (split-half $R^2$: \ci{0.91}{0.89}{0.93}), a measure of reliability in our item-level estimates of performance.
The relative size of this gap between even the best performing three-factor model and the noise ceiling suggests that the majority of the variance to be explained requires a model that can capture more subtle characteristics of each item that cause some to be more challenging than others. 


\begin{figure}[t!]
    \centering 
    \includegraphics[width=0.75\columnwidth]{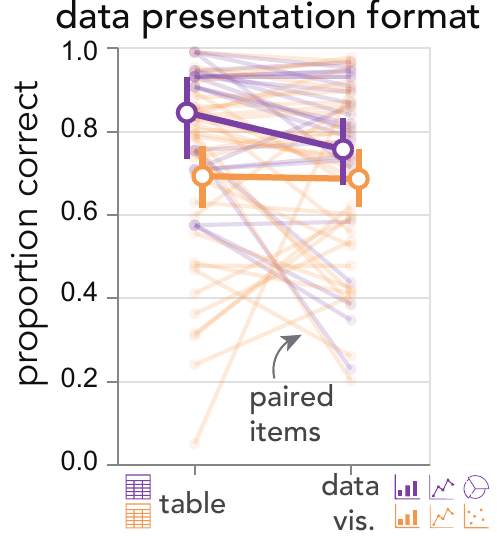}
    \caption{Performance across different \emph{data presentation formats} (i.e. table or data visualization). Each line segment represents a pair of items in either \wanier{} and \brbf{} which differ only in \emph{data presentation format}. Error bars represent bootstrapped 95\% confidence intervals.}
    \label{fig:performance_by_presentation}
    \vspace{-1em}
\end{figure}

\vspace{-1em}
\paragraph{How does performance vary between different \emph{data presentation formats}?}

One potential barrier to answering certain questions about data visualizations might be the level of precision provided by these displays by comparison with other formats for displaying quantitative information, such as tables. 
For example, a \texttt{value identification} question might even be easier to answer precisely with a \texttt{table} than with a \texttt{scatter plot}.
To explore potential differences in difficulty attributable to \emph{data presentation format}, we included several pairs of items from two of the tests \citep{wainer1980test, boy2014principled} that were otherwise identical, except that one of them included a \texttt{table} and the other a data visualization.

We did not find evidence for an overall difference in performance on items with tables and those with data visualizations ($t(59)$~=~-1.42, $p$~=~.160), neither in \brbf{} ($\Delta$table $-$ data~visualization: \ci{0.01}{ -0.10}{0.10}) nor in \wanier{} ($\Delta$ table $-$ data~visualization: \ci{0.09}{-0.04}{0.22}; Figure~\ref{fig:performance_by_presentation}).
These null results are compatible with several possibilities, including that data presentation formats do not strongly impact performance and the suite of items included in this evaluation were insufficient to resolve a global difference between formats.  
Given the substantial variability we observed across pairs of items, however, it seems more plausible that the degree to which data presentation format impacts performance depends more strongly on other factors (e.g., the question being asked, other characteristics of the data). 


\vspace{-0.5em}
\section{Discussion}
\vspace{-0.5em}
Here we administered five assessments of data visualization literacy to a large sample of U.S. adults to obtain precise estimates of the difficulty of these items under consistent test conditions.
We found that various features of these items (i.e., \emph{graph type}, \emph{task type}) could explain some item-level variation in performance, but there was substantial and reliable variation left unexplained.
Thus, other features, or additional features, are needed in order to predict why some of them are more difficult than others. 

In recent years, there has been broadening agreement on a general conceptual framework for data visualization literacy \citep{borner2019data, hedayati2024pixels}. 
However, there is not yet a consensus on a concrete set of instruments for measuring these literacy skills in a comprehensive manner, organized around components that are predictive of detailed patterns of performance.
Our findings suggest major opportunities to develop unified measures of data visualization literacy which reliably evaluate the same skill set across individuals.


Having more unified measures is especially valuable for contributing to development and evaluation of computational models that can make explicit predictions on an item-by-item basis, as well as be used to test more specific hypotheses concerning the underlying mechanisms that support the understanding of complex visual inputs. 
In the future, such models might have the potential to explain in mechanistic terms why some operations with visual displays are more difficult for people than others \citep{nobre2024reading}, and what strategies might be useful for overcoming those barriers. 
In addition, improved measures and models might help to account for reliable variation in data visualization literacy across individuals, who might have varying amounts of prior experience with mathematics, statistics, and other data-intensive subjects. 
In the long run, improved understanding of the mechanisms that support data visualization understanding might be leveraged to develop improved ways of helping more people acquire core quantitative literacy skills.

\section{Acknowledgments}
We thank members of the Cognitive Tools Lab for their input and feedback. This work was supported by NSF CAREER Award \#2436199, NSF DRL \#2400471, and awards from ONR Science of Autonomy, the Stanford Human-Centered AI Institute (HAI) and Stanford Accelerator for Learning.

\section{Data and code availability}
All experimental materials, data, and analysis code are available at \url{https://github.com/cogtoolslab/viz_item_measures_cogsci2025}.


\bibliographystyle{apacite}
\setlength{\bibleftmargin}{.125in}
\setlength{\bibindent}{-\bibleftmargin}
\bibliography{bibliography}

\end{document}